\documentclass{article}

\usepackage{arxiv}

\usepackage[utf8]{inputenc} 
\usepackage[T1]{fontenc}    
\usepackage{hyperref}       
\usepackage{url}            
\usepackage{booktabs}       
\usepackage{amsfonts}       
\usepackage{nicefrac}       
\usepackage{microtype}      
\usepackage{lipsum}		
\usepackage{graphicx}
\usepackage{natbib}
\usepackage{doi}
\usepackage{bm}
\usepackage[varvw]{newtxmath}       
\usepackage{amsmath}
\mathchardef\mhyphen="2D 
\usepackage{booktabs}
\usepackage{multirow}
\usepackage{comment}

\title{Uncertainty Quantification and Sensitivity analysis for Digital Twin Enabling Technology: Application for BISON Fuel Performance Code}

\author{ {Kazuma ~Kobayashi}\\
	Department of Nuclear Engineering and Radiation Science\\
	Missouri University of Science and Technology\\
	Rolla, MO 65409, USA \\
	\And
	{Dinesh ~Kumar } \\
	Department of Mechanical Engineering\\
	University of Bristol\\
	Bristol BS8 1TR, UK \\
\And
	{Matthew ~Bonney} \\
	Department of Mechanical Engineering\\
	University of Sheffield\\
	Sheffield S10 2TN, UK\\
 \And
 	{Souvik  ~Chakraborty} \\
	Department of Applied Mechanics\\
	School of Artificial Intelligence\\
        Indian Institute of Technology Delhi\\
	Hauz Khas - 110016, New Delhi, India\\
 \And
 	{Kyle  ~Paaren} \\
	Fuel Development, Performance, and Qualification\\
	Idaho National Laboratory\\
       Idaho Falls, ID  83415  \\
 \And
	{Syed ~Alam} \\
	Department of Nuclear Engineering and Radiation Science\\
	Missouri University of Science and Technology\\
	Rolla, MO 65409, USA \\
 }

\renewcommand{\shorttitle}{\textit{Handbook of Smart Energy Systems} }

\hypersetup{
pdftitle={A template for the arxiv style},
pdfsubject={q-bio.NC, q-bio.QM},
pdfauthor={David S.~Hippocampus, Elias D.~Striatum},
pdfkeywords={First keyword, Second keyword, More},
}

\begin{document}
\maketitle

\begin{abstract}
To understand the potential of intelligent confirmatory tools, the U.S. Nuclear Regulatory Committee (NRC) initiated a future-focused research project  to assess the regulatory viability of machine learning (ML) and artificial intelligence (AI)-driven  Digital Twins (DTs) for nuclear power applications. Advanced accident tolerant fuel (ATF) is one of the priority focus areas of the U.S. Department of Energy (DOE). A DT framework can offer game-changing yet practical and informed solutions to the complex problem of qualifying advanced ATFs. Considering the regulatory standpoint of the modeling and simulation (M\&S) aspect of DT, uncertainty quantification and sensitivity analysis are paramount to the DT framework's success in terms of multi-criteria and risk-informed decision-making. This chapter introduces the ML-based uncertainty quantification and sensitivity analysis methods while exhibiting  actual applications to the finite element-based nuclear fuel performance code BISON.
\end{abstract}


\section{Introduction}
The increasing performance of computers has made tremendous contributions to the industrial field. In particular, large industrial products such as power plants require a great deal of time and money just for typical prototyping. To reduce overall cost, time, and materials, a need for computational simulations has risen. One recent advancement that has garnered industrial attention is the digital twin (DT) concept, which can advance the design and prototyping aspects of the system and the manufacturing and asset management throughout the system's life cycle. The DT technology is expected to significantly contribute as a higher-level decision-making factor from the design stage, especially when creating products and structures that could have a significant human, financial, and environmental impact during an accident, i.e., nuclear power plants \cite{kobayashi2022practical, kobayashi2022digital, rahman2022leveraging}. In designing a DT-enabling technology, it is indispensable to have the necessary numerical models such as optimization algorithms, and it is undoubtedly the most important to prepare a suitable model. 

Since the Fukushima accident in 2011, the US Department of Energy (DOE) has been developing nuclear fuel and cladding candidates with increased accident tolerance under the "ATF Program" \cite{osti_doe}. Through the DOE project, potential candidate materials have been narrowed. There are many universities, national laboratories, and nuclear vendors involving the R\&D of ATF. However, their long-term perspectives for cladding focus on Silicon-fiber reinforced SiC-matrix composites ($\rm SiC_{f}/SiC_{m}$). Compared to Zr-based cladding, silicon carbide-based ceramics have better neutron economy and chemical inertness at high temperatures \cite{singh2018parametric}. Moreover, $\rm SiC_{f}/SiC_{m}$ are considered as an appropriate material for light water reactors (LWRs) in terms of its superior properties: stability under irradiation, fracture toughness, and oxidation resistance to high-temperature steam up to 1700 $\rm ^{\circ}C$ \cite{singh2018thermo}. Although there are other candidates such as Cr-coated Zr-based alloy or FeCrAl, their properties are outperformed by $\rm SiC_{f}/SiC_{m}$. There is also a weakness; SiC does not want to creep, allowing stress to build up further as fuel-cladding mechanical interaction (FCMI) continues at higher burnup, leading to more fuel failures. Based on these facts, the nuclear industry has shown great interest in using $\rm SiC_{f}/SiC_{m}$ but faces the problem of a statistical deficiency (or lack of data). In other words, material property data is uncertain. In particular, density and thermal properties influence neutronics and heat transfer calculations during nuclear fuel system design. Considering the inherent benefit of DT as a decision-making framework, there are significant efforts from the NRC to initiate future applications of DT for conforming robust decisions on Accident Tolerant Fuel (ATF) in nuclear safety and risk analysis and satisfying NRC's regulatory requirements.

The major  challenges related to M\&S for DT are (a) Model integration through coupling, (b) Incorporating ML/AI algorithms, (c) Treatment of noisy or erroneous data, (d) Lack of data, and missing data, and (e) Uncertainty quantification and sensitivity. The method of quantitatively evaluating the impact of uncertain input variables on system output is called uncertainty quantification (UQ), and sensitivity analysis (SA) is the analysis of the contribution from each input variable to its output. Although there are several UQ methods, this chapter focuses on the Polynomial Chaos Expansion (PCE). PCE is more computationally efficient than more classical methods such as Monte Carlo (MC), or interval analysis \cite{Kumar2020, Kumar2021}. These works demonstrate this for the nuclear field by performing the PCE method using a finite element-based nuclear fuel performance code BISON \cite{LaboratoryIdahoNational}. This chapter will incorporate UQ and SA into the DOE fuel performance code BISON for future integration into the DT framework.

\section{Polynomial Chaos for Uncertainty Quantification}

The polynomial chaos method (PCM) is a UQ technique with great potential for stochastic simulations. Mean, variance, higher-order moments, and the probability density function can be used to describe the properties of the input stochastic variables and the output stochastic solution \cite{Kumar2016, Kumar2020a}.

Orthogonal polynomials are polynomial classes that are orthogonal to one another in terms of a weight function. Popular orthogonal polynomials utilized in PC-based stochastic applications include Hermite polynomials, Laguerre polynomials, Jacobi polynomials, and Legendre polynomials. Orthogonality means:
\begin{equation}
    \int_{\xi}\psi_{i}(\xi)\psi_{j}(\xi)W_{\xi}(\xi)d\xi = \left< \psi_{i}\psi_{j} \right> = \delta_{ij} \left< \psi_{i}^{2} \right>
\end{equation}
where $W_{\xi}(\xi)$ is the probability distribution of the random variable $\xi$, $\delta_{ij}$ is the Kronecker delta, $\psi_{i}(\xi)$ are basis functions, and $\left< \psi_{i}\psi_{j} \right>$ represents the inner product.

It is possible to separate deterministic and non-deterministic orthogonal polynomials in PCMs. For instance, the decomposition of a random variable $u(x, \xi)$ is:
\begin{equation}
\label{eq:decompose}
    u(x, \xi) = \sum_{i=0}^{P} u_{i}(x)\psi_{i}(\xi)
\end{equation}
where $u_{i}(x)$ are the deterministic expansion coefficients and  $P$ is the total number of terms in the expansion.

The mean of $u(x)$ can be written as:
\begin{equation}
\label{eq:mean}
    E[u] = u_{0}
\end{equation}

and the variance as:
\begin{equation}
\label{eq:var}
    E[(u- E[u])^2] = \sigma_{u}^{2} = \sum_{i=1}^{P}u_{i}^{2} \left< \psi_{i}^{2} \right>
\end{equation}

Typically there are 20+ unknown parameters going into the simulations and because of the PCE replaces each parameter with a set of unknown parameters (say 6 per thus resulting in 120 parameters). Since high-dimensional stochastic problems require exponentially more computation, this is known as the "curse of dimensionality" and is the primary drawback of all PCM methods. For this reason, the development of efficient stochastic models for the analysis of uncertainty in complex industrial applications is of great interest \cite{Kumar2016}.

\subsection{Multi-dimensional or multivariable polynomials}
Building multi-dimensional polynomials from one-dimensional ones is necessary to study stochastic variables' impact on the final solution. The multi-dimensional PCE of order $p$ can be expressed in terms of 1D polynomials. For demonstration, a 2D PCE of order $p=3$ is chosen.

When the set of 1-dimensional orthogonal polynomials of PC order 3 is defined as $\{ \Psi_{0},\Psi_{1}, \Psi_{2}, \Psi_{3} \}$, a 2D stochastic quantity $u(\xi_{1}, \xi_{2})$ is expressed as a combination of the polynomial functions and coefficients.

\begin{equation}
    \begin{split}
        u(\xi_{1},\xi_{2}) &= u_{00}\Psi_{0}\\
        &+ u_{10}\Psi_{1}(\xi_{1}) + u_{01}\Psi_{1}(\xi_{2})\\
        &+ u_{20}\Psi_{2}(\xi_{1}) + u_{11}\Psi_{1}(\xi_{1})\Psi_{2}(\xi_{2})  + u_{02}\Psi_{2}(\xi_{2})\\
        &+ u_{30}\Psi_{3}(\xi_{1}) + u_{21}\Psi_{2}(\xi_{1})\Psi_{1}(\xi_{2}) + u_{12}\Psi_{1}(\xi_{1})\Psi_{2}(\xi_{2}) + u_{03}\Psi_{3}(\xi_{2}).
    \end{split}
\end{equation}

For a set of multi-dimensional independent variable, $\boldsymbol{\xi} = (\xi_{1},...,\xi_{n})$, the probability density function (PDF) can be define by the following:

\begin{equation}
\label{eq:weight}
    \boldsymbol{W}(\boldsymbol{\xi}) = \prod_{i=1}^{n}W_{i}(\xi_i)
\end{equation}
where $W_{i}(\xi_i)$ is the individual PDF of the random variable $\xi_{i}$ \cite{Kumar2016, Kumar2020a}. The sum of the polynomial terms $P+1$ in Eq. \ref{eq:decompose} can be expressed with the polynomial order ($p$) and number of input variables ($n$) as:

\begin{equation}
\label{eq:tot_pols}
    P+1 = \frac{(p+n)!}{p!n!}
\end{equation}
Therefore, the sum of the polynomial terms ($P+1=10$) can be computed for the demonstration ($p=3, n=2$).

\subsection{Regression method to estimate PC coefficients}
A common way to determine the unknown set of polynomial coefficients, a regression analysis is performed. Regression analysis is a set of statistical procedures used to estimate the relationship between a dependent variable and one or more independent variables in statistical modeling. In the most typical type of regression analysis, known as linear regression, the line that most closely matches the data in terms of a given mathematical criterion is found. This method is often used in data analysis and in surrogate modeling methods such as Gaussian processes \cite{kobayashi2022practical}.

Walters' regression-based non-intrusive polynomial chaos method computes polynomial coefficients \cite{Kumar2020a, Walters2003}. In the sampling-based regression method, the unknown variables are written as the polynomial expansions of those variables. The approximate PCE for the stochastic quantity of interest, $u(x;\boldsymbol{\xi})$, is as follows:

\begin{equation}
    u(x;\boldsymbol{\xi}) = \sum_{i=0}^{P}u_{i}(x)\psi_{i}(\boldsymbol{\xi})
\end{equation}

Using $m$ samples $(\boldsymbol{\xi^{j}}= \{ \xi_{1},...,\xi_{n_{s}} \}^{j}; j=1,...,m)$ from the PDF $\mathbf{W}(\mathbf{\xi})$ from Eq. \ref{eq:weight} and the corresponding model output  $u(x;\boldsymbol{\xi}^{j})$, this system is described as a matrix equation:

\begin{equation}
\begin{bmatrix}
\psi_{0}(\boldsymbol{\xi}^{1}) && \psi_{1}(\boldsymbol{\xi}^{1}) && \cdots && \psi_{P}(\boldsymbol{\xi}^{1}) \\
\psi_{0}(\boldsymbol{\xi}^{2}) && \psi_{1}(\boldsymbol{\xi}^{2}) && \cdots && \psi_{P}(\boldsymbol{\xi}^{2}) \\
\vdots  && \vdots && && \vdots\\
\psi_{0}(\boldsymbol{\xi}^{m}) && \psi_{1}(\boldsymbol{\xi}^{m}) && \cdots && \psi_{P}(\boldsymbol{\xi}^{m}) \\
\end{bmatrix}
\begin{bmatrix}
 {u_{0}}(x) \\ {u_{1}}(x) \\ \vdots \\ {u_{P}}(x)
\end{bmatrix}
=
    \begin{bmatrix}
u(x;\boldsymbol{\xi}^{1}) \\ u(x;\boldsymbol{\xi}^{2}) \\ \vdots \\ u(x;\boldsymbol{\xi}^{m})
    \end{bmatrix}
\end{equation}

or in matrix notation

\begin{equation}
\label{eq:system}
    [A]\{u\} = \{b\}
\end{equation}

The objective is to solve this equation for $\{u\}$; it can be expressed using matrix notation with the assumption of $m>P$:

\begin{equation}
\label{eq:f_matrix}
    \{u\} = \left( [A]^{\top}[A] \right)^{-1}[A]^{\top}\{b\}
\end{equation}

By combining Eqs. \ref{eq:mean}, \ref{eq:var}, \ref{eq:weight} and \ref{eq:f_matrix}, the mean and variance of system outputs can be quantitatively evaluated by considering uncertain input variables.

\section{Sensitivity Analysis}
Although the system response due to uncertain input variables is quantified with the UQ methods, the contribution from the individual inputs should be considered. Sobol sensitivity is one of the global sensitivity analysis methods \cite{sobol2001global}. In this method, the system response, $u(x)$, is decomposed associating with the input $x=(x_{1},\dots,x_{n})$:

\begin{equation}
\label{eq:sa_base}
u(x) = u_{0} + \sum_{s=1}^{n}\sum_{i_{1},\cdots,i_{s}}^{n} u_{i_{1} \dots i_{s}}(x_{i_{1}},\cdots,x_{i_{s}})   
\end{equation}
\noindent
where $i \leq i_{1} < \dots < i_{s} \leq n$. Also, Eq. \ref{eq:sa_base} can be expressed with analysis of variance (ANOVA) \cite{cardinal2013anova} representation,

\begin{equation}
\label{eq:anova}
    u(x) = u_{0} + \sum_{i}u_{i}(x_{i}) + \sum_{i<j}u_{ij}(x_{i},x_{j}) + \dots + u_{12 \dots n}(x_{1}, x_{2},\dots,x_{n}).
\end{equation}
\noindent
under the condition of
\begin{equation}
    \int_{0}^{1}u_{i_{i},\dots,i_{s}}(x_{i_{1}},\dots,x_{i_{s}})dx_{k} = 0
\end{equation}
\noindent
where $k=i_{1},\cdots,i_{s}$.

The individual term can be analytically described as:
\begin{equation}
    \int u(x)dx = u_{0},
\end{equation}

\begin{equation}
    \int u(x) \prod_{k\neq i}dx_{k} = u_{0} + u_{i}(x_{i}),
\end{equation} 

\begin{equation}
    \int u(x) \prod_{k \neq i,j}dx_{k} = u_{0} + f_{i}(x_{i}) + f_{j}(x_{j}) + f_{ij}(x_{i},x_{j}).
\end{equation}

Based on these descriptions, Sobol \cite{sobol2001global} introduced a method for calculating the variance of a response function, $u(x)$, by assuming it to be square integrable. The variances are expressed as:

\begin{equation}
    V = \int u^{2}(x)dx - u_{0}^{2}
\end{equation}
\noindent
and
\begin{equation}
    V_{i_{1} \dots i_{s}} = \int u_{i_{1} \dots i_{s}}^{2}dx_{i_{1}} \dots dx_{i_{s}}.
\end{equation}
\noindent
The relationship between the above variances are

\begin{equation}
    V = \sum_{s=1}^{n}\sum_{i_{1}<\dots<i_{s}}^{n}V_{i_{1} \dots i_{s}}.
\end{equation}
Finally, the ratios of variances are called Sobol indices.

\begin{equation}
    S_{i_{1}\dots i_{s}} = \frac{ V_{i_{1} \dots i_{s}}}{V}
\end{equation}
where $S_{i}$ is a measure of the first order sensitivity express the effect of the total variance originating from the uncertainties in the set of input variable $x$. It is known that the Sobol indices can be acquired from the PCE \cite{Kumar2020}. Therefore, a set of the PCEs and the system response are acquired, and the Sobol indices can be computed.

\section{Demonstrations to Fuel Performance Code: BISON}

In order to demonstrate the uncertainty quantification method and sensitivity analysis to the BISON code, a model with simple geometry composed of nuclear fuel and cladding material properties is used.

\begin{figure}[htbp]
    \centering
    \includegraphics[width=4cm]{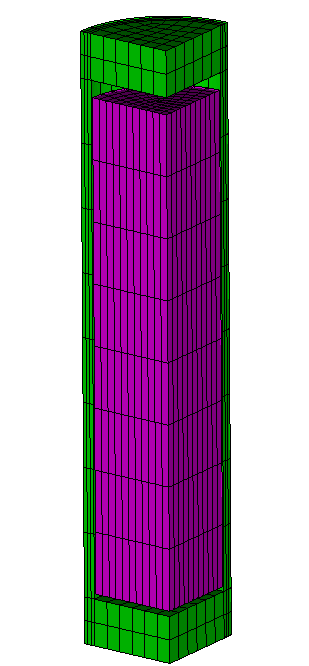}
    \caption{3D mesh of a fuel pin. A purple region represents a fuel domain, and a green one shows a cladding. }
    \label{fig:geometry}
\end{figure}

\begin{figure}[htbp]
    \centering
    \includegraphics[width=15cm]{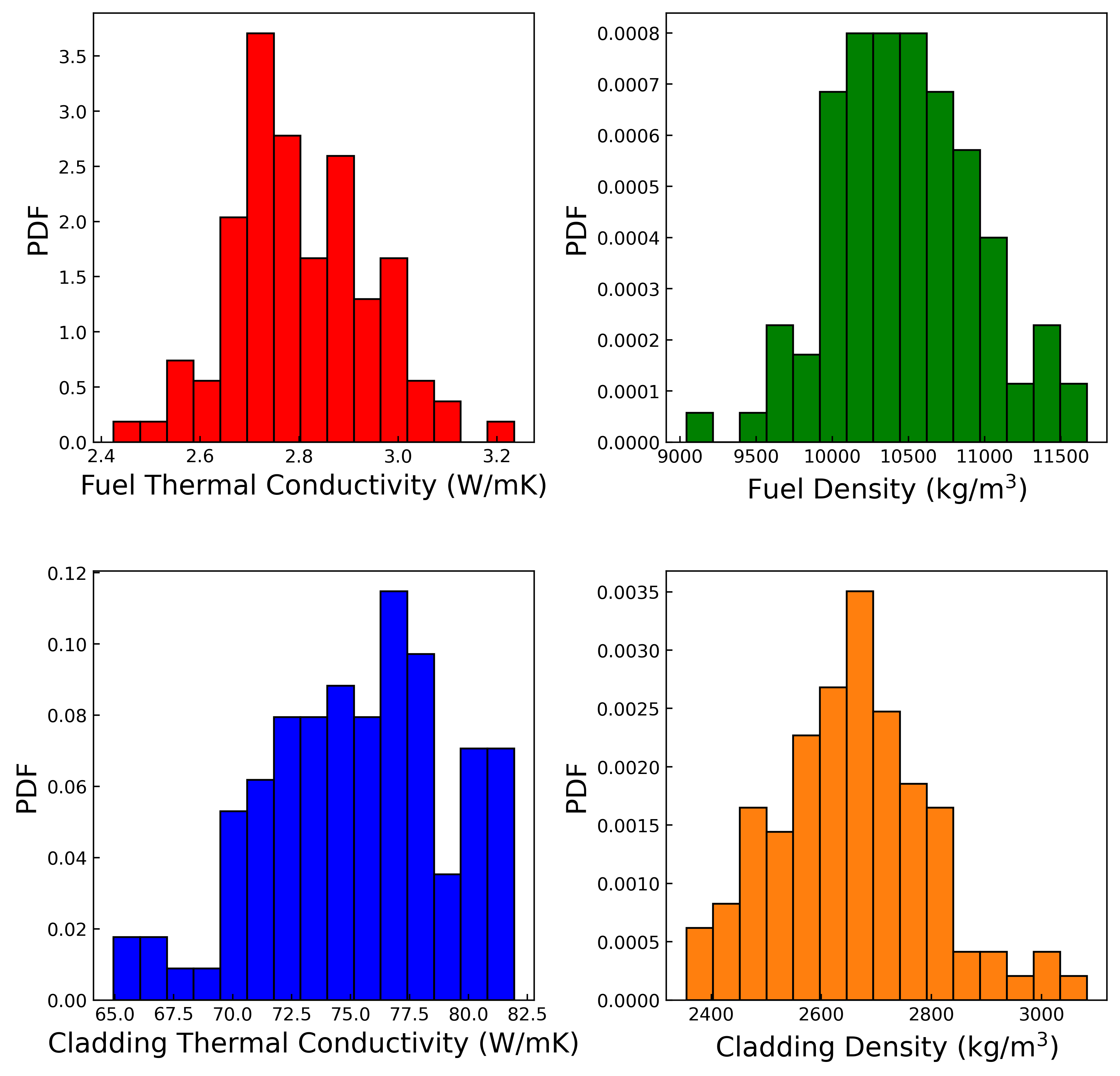}
    \caption{Probability distributions of input variables}
    \label{fig:input_pdf}
\end{figure}

Fig. \ref{fig:geometry} represents the geometry, and the geometric design parameters are listed in Table \ref{tab:geometry}. In this demonstration, uncertain input variables were room temperature values of thermal conductivity and mass density of $\rm UO_{2}$ fuel and cladding as listed in Table \ref{tb:para_input}. All input variables are assumed to follow a normal distribution with a 5\% coefficient of variation, which is standard deviation/mean, and the MC sampling method is employed. The number of samples needs to be more than 35 from Eq. \ref{eq:tot_pols}, therefore, 100 samples are generated to ensure the over-sampling. The probability distribution function (PDF) for each variables is shown in Fig. \ref{fig:input_pdf}.

\begin{table}[htbp]
\centering
\caption{Geometric parameters of fuel, gap, and cladding}
\label{tab:geometry}
\begin{tabular}{@{}llllll@{}}
\toprule
\textbf{Domain Name}      &  & \textbf{Parameters} &  &  & \textbf{Values (mm)} \\ \midrule
\multirow{2}{*}{(1) Fuel}     &  & radius              &  &  & 4.1                  \\
                          &  & height              &  &  & 26.2                 \\ \midrule
\multirow{3}{*}{(2) Gap}      &  & plenum              &  &  & 0.817                \\
                          &  & radial              &  &  & 0.123                \\
                          &  & axial               &  &  & 0.25                 \\ \midrule
\multirow{5}{*}{(3) Cladding} &  & inner radius        &  &  & 4.22                 \\
                          &  & outer radius        &  &  & 4.74                 \\
                          &  & radial thickness    &  &  & 0.517                \\
                          &  & axial thickness     &  &  & 2.24                 \\
                          &  & height              &  &  & 29.3                 \\ \bottomrule
\end{tabular}
\end{table}

\begin{table}[htbp]
\caption{List of input parameters for BISON}
\label{tb:para_input}
\centering
\begin{tabular}{@{}lll@{}}
\toprule
\textbf{Input variables}      & \textbf{Mean} & \textbf{Std.} \\ \midrule
Fuel thermal conductivity $\rm (W/mK)$ \cite{wu2022mechanism}    & 2.8           & 0.1           \\
Fuel density $\rm (kg/m^{3})$ \cite{wu2022mechanism}                 & 10430.0       & 521.5         \\
Clad thermal conductivity $\rm (W/mK)$ \cite{Kowbel} & 75.0          & 3.8           \\
Clad density $\rm (kg/m^{3})$ \cite{Kowbel}           & 2650.0        & 132.5         \\ \bottomrule
\end{tabular}
\end{table}

\begin{figure}[htbp]
    \centering
    \includegraphics[width=15cm]{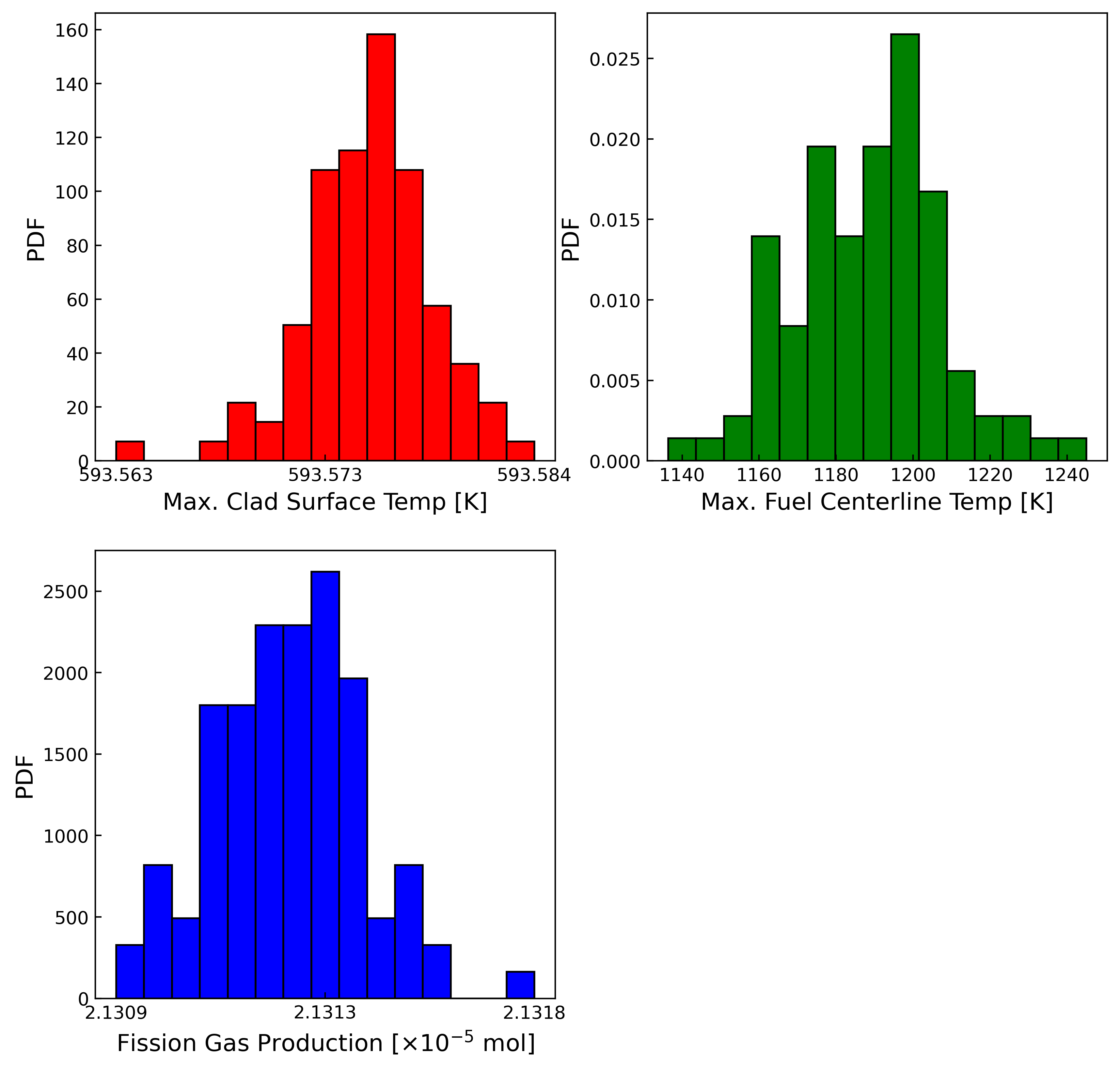}
    \caption{Probability distributions of output responses at 176.8 days.}
    \label{fig:output_pdf}
\end{figure}

The outputs were maximum cladding surface temperatures, maximum fuel centerline temperatures, and fission gas production in the fuel. The time from before the reactor operation to one year later was divided into 29 steps, and the outputs at each step were obtained. For example, Fig. \ref{fig:output_pdf} represents the PDFs at 176.8 days after the reactor operation started. 

\begin{figure}[htbp]
    \centering
    \includegraphics[width=15cm]{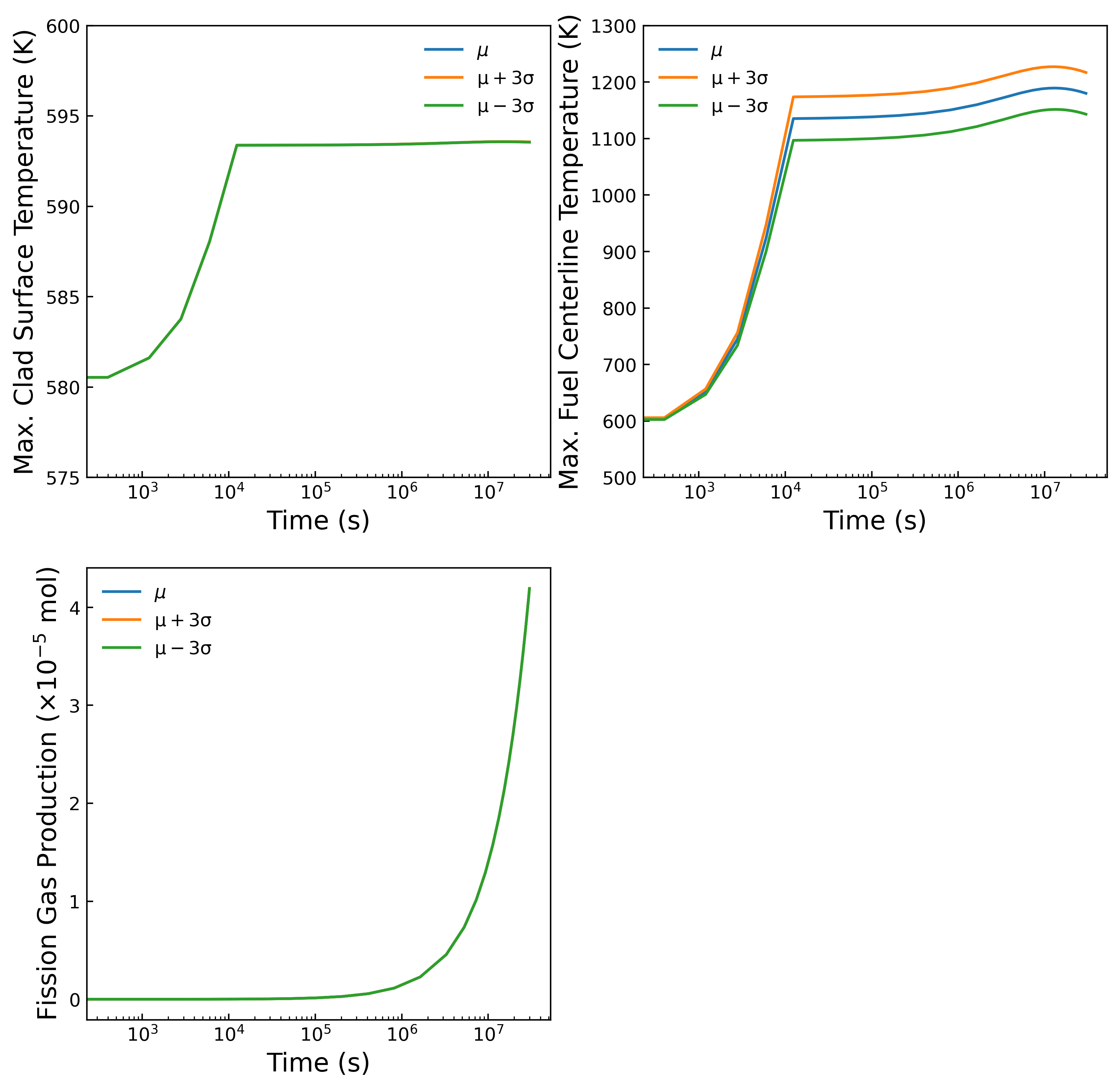}
    \caption{Results of UQ calculation (mean and uncertainty).}
    \label{fig:uq}
\end{figure}

The time evolution of the PDFs produced from the PCE procedure is shown in Fig. \ref{fig:uq}. The x-axis corresponds to the reactor operation time in seconds and the y-axis the output value. It was provided by solving Eq. \ref{eq:f_matrix}. The means and the variances for each output value were introduced by Eq. \ref{eq:mean} and \ref{eq:var}, respectively. The blue solid line shows the mean values, and the orange and green ones represent variances ($3\sigma$). It was observed that the maximum cladding surface and fuel centerline temperatures are proportional to the reactor operation time by $\sim 10^{4}$ seconds, and then keep an almost flat value with slight fluctuations. In contrast, the fission gas production is prolonged and starts a quick increment after $\sim 10^{6}$ seconds. For the maximum cladding surface temperatures and fission gas productions, it can be seen that the uncertainties due to the input variables are not critical. However, these results show the case with a relatively small amount of uncertainty. Therefore, when the uncertainty of input is expected to be greater than that value, the UQ calculation must be performed again for new input parameters.

\begin{figure}[htbp]
    \centering
    \includegraphics[width=15cm]{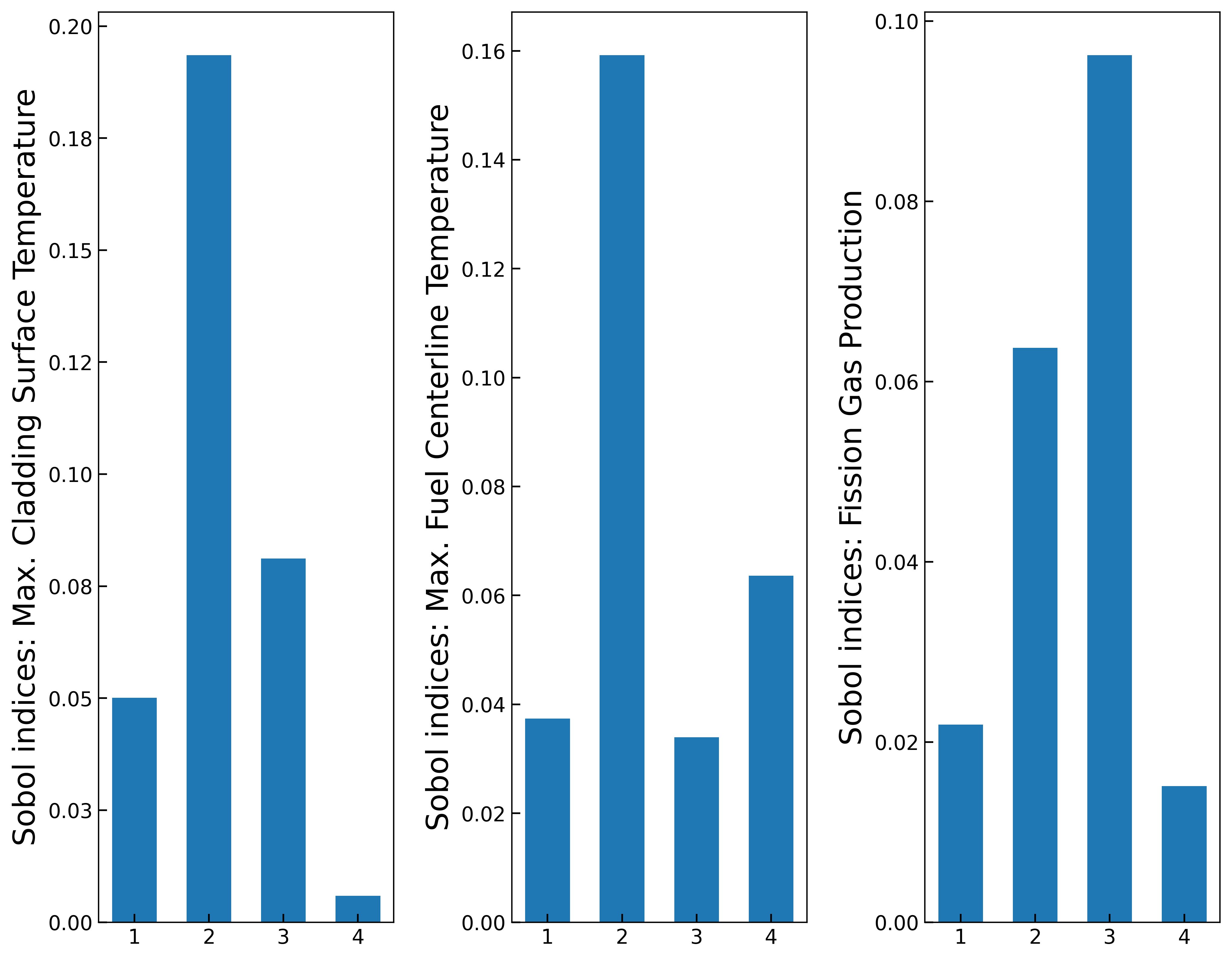}
    \caption{Sensitivity analysis: Sobol indices of the outputs with respect to the input uncertainties, 1: fuel thermal conductivity, 2: fuel density, 3: cladding thermal conductivity, and 4: cladding density.}
    \label{fig:sa}
\end{figure}

The sensitivity analysis was performed with the results of uncertainty quantification. The impact of uncertain inputs on output responses was quantified and the results are shown in Fig. \ref{fig:sa}. The x-axis shows the input variable name, 1: fuel thermal conductivity, 2: fuel density, 3: cladding thermal conductivity, and 4: cladding density. The y-axis shows the Sobol indices of the outputs with respect to the input uncertainties. The analysis reveals the relationships between input variables and output responses:

\begin{itemize}
    \item {Cladding surface temperature: the largest contribution comes from the fuel density and the smallest one from the cladding density}
    \item {Fuel centerline temperature: the largest contribution comes from the fuel density and the smallest one from the cladding thermal conductivity}
    \item {Fission gas production: both the fuel density and cladding thermal conductivity have large contributions, and the smallest contribution comes from the cladding density}
\end{itemize}

\section*{Conclusion}
The use of digital twin technology is a top priority for future nuclear power development, data analysis and modeling methods that apply artificial intelligence and machine learning are expected to develop further in the future. Among Machine learning-based data-driven methods, uncertainty quantification and sensitivity analysis are necessary in terms of designing complex systems such as nuclear reactors. However, their applications in nuclear power systems are very limited right now. The demonstration was performed with one of the next-generation nuclear fuel performance codes to introduce these methods to the nuclear field: BISON. As a result, the impact of uncertain inputs and their contribution to the output values were quantified. An advantage of these methods is to treat a simulator as a black box. In other words, the user can implement these methods into any other nuclear codes: for example, neutronics, CFD, fuel performance, and even their coupling simulations. This advantage will significantly benefit the evaluation of new materials for advanced reactor systems (e.g. accident tolerant fuels, high-temperature gas reactors, and fusion reactors), with significant uncertainties due to a lack of experimental data. In another application, it is possible to evaluate the impact of each reactor design parameter on the entire system and quantify the impact on the operation. In conclusion, uncertainty quantification and sensitivity analysis may contribute to the nuclear field: by accelerating R\&D and enhancing risk assessment.

\section*{Acknowledgement}
The computational part of this work was supported in part by the National Science Foundation (NSF) under Grant No. OAC-1919789.

\bibliographystyle{unsrtnat}
\bibliography{references}  






\end{document}